\documentclass[12pt]{article}
\usepackage{graphicx}
\usepackage{amssymb}
\usepackage{amsmath}
\numberwithin{equation}{section}

\begin{document}
\author{Lev Sakhnovich}
\date{August15, 2006}
\textbf{Rational solutions of Knizhnik-Zamolodchikov system}

\begin{center} Lev Sakhnovich  \end{center}
735 Crawford ave., Brooklyn, 11223, New York, USA.\\
 E-mail address:lev.sakhnovich@verizon.net
\begin{center}Abstract \end{center}
We consider  Knizhnik-Zamolodchikov system of linear
differential equations. The coefficients of this system are
rational functions. We prove that under some conditions the
solution of KZ system is rational too. This assertion confirms
partially the conjecture of Chervov-Talalaev.\\
\textbf{Mathematics Subject Classification (2000).} Primary 34M05,
Secondary 34M55, 47B38.\\
\textbf{Keywords.} Symmetric group, natural representation, linear
differential system, rational fundamental solution.

\section{Main Theorem}
1.We consider the differential system
\begin{equation}
\frac{dW}{dz}={\rho}A(z)W, \quad z{\in}C,\end{equation} where
$\rho$ is integer, $A(z)$ and $W(z)$ are $n{\times}n$ matrix
functions. We suppose that $A(z)$ has the form
\begin{equation}
A(z)=\sum_{k=1}^{s}\frac{P_{k}}{z-z_{k}},\end{equation} where
$z_{k}{\ne}z_{\ell}$ if $k{\ne}\ell$. In a neighborhood of $z_{k}$
the matrix function $A(z)$
 can be represented in the form
\begin{equation}
A(z)=\frac{a_{-1}}{z-z_{k}}+a_{0}+a_{1}(z-z_{k})+... ,
\end{equation}where $a_{k}$ are $n{\times}n$ matrices.
We investigate the case when $z_{k}$ is either a regular point of
$W(z)$ or a pole. Hence the following relation
\begin{equation}
W(z)=\sum_{p{\geq}m}b_{p}(z-z_{k})^{p},\quad b_{m}{\ne}0
\end{equation} is true. Here $b_{p}$ are $n{\times}n$ matrices.
We note that $m$ can be negative. \\ \textbf{Proposition
1.1.}(necessary condition, (see \textbf{[Sa06]})  \emph{If the
solution of system $(1.1)$ has form $(1.4)$ then $m$ is an
eigenvalue of
$a_{-1}$.}\\
We denote by M the greatest integer eigenvalue of the matrix
${\rho}a_{-1}$. Using relations ((1.3) and (1.4)) we obtain the assertion.\\
\textbf{Proposition 1.2.}(necessary and sufficient condition, (see
\textbf{[Sa06]}) \emph{If the matrix system
\begin{equation}
[(q+1)I_{n}-a_{-1}]b_{q+1}=\sum_{j+\ell=q}{\rho}a_{j}b_{\ell},
\end{equation} where $m{\leq}q+1{\leq}M$ has a solution $b_{m},
b_{m+1},...,b_{M}$ and $b_{m}{\ne}0$ then system (1.1) has a
solution
of form (1.4).}\\
We introduce the matrices
\begin{equation}
P_{k}^{-}=I+P_{k},\quad P_{k}^{+}=I-P_{k}.\end{equation}
\textbf{Theorem 1.1.} \emph{Let the following conditions be
fulfilled}\\
1)\begin{equation} P_{k}^{2}=I_{n},\quad
1{\leq}k{\leq}s.\end{equation}
2)\begin{equation}[P_{j}P_{k}P_{\ell}+P_{\ell}P_{k}P_{j}]P_{k}^{+}=0,\quad
j{\ne}k,\quad j{\ne}\ell,\quad k{\ne}\ell.\end{equation} 3)
\begin{equation}(P_{j}P_{k}P_{j}+P_{j})P_{k}^{+}=P_{k}^{+},\quad
j{\ne}k.\end{equation} 4) The matrices $P_{j}\quad
(1{\leq}j{\leq}s)$ are symmetric.\\
\emph{If  $\rho={\pm}1$ then system $(1.1)$, $(1.2)$ has
a rational fundamental solution .} \\
\emph{Proof.} We shall use the notations  (1.2) and (1.3). Then in
neighborhood of $z_{k}$ we have
\begin{equation}
a_{-1}=P_{k},\quad
a_{r}=(-1)^{r}\sum_{j{\ne}k}\frac{P_{j}}{(z_{k}-z_{j})^{r+1}},\quad
r{\geq}0.\end{equation} Using relation (1.5) and equality
\begin{equation}P_{k}^{-}P_{k}^{+}=0 \end{equation} we have
\begin{equation}
b_{-1}=P_{k}^{+},\quad
b_{0}=-\sum_{j{\ne}k}P_{k}\frac{P_{j}}{(z_{k}-z_{j})}b_{-1}.\end{equation}
Formulas (1.5) and (1.12) imply that \begin{equation}
P_{k}^{+}b_{1}=-[\sum_{j{\ne}k}\frac{P_{j}}{z_{k}-z_{j}}\sum_{\ell{\ne}k}P_{k}\frac{P_{\ell}}{z_{k}-z_{\ell}}
+\sum_{j{\ne}k}\frac{P_{j}}{(z_{k}-z_{j})^{2}}]b_{-1}.\end{equation}
Due to conditions (1.8) and (1.9) we can write (1.12) in the form
\begin{equation}P_{k}^{+}b_{1}=-\sum_{j{\ne}k}\frac{1}{(z_{k}-z_{j})^{2}}P_{k}^{+}.\end{equation}
Equation (1.14) has the following solution
\begin{equation}
b_{1}=-\beta_{k}I_{n},\quad when \quad
\beta_{k}=\sum_{j{\ne}k}\frac{1}{(z_{k}-z_{j})^{2}}{\ne}0\end{equation}
and \begin{equation}b_{1}=P_{k}^{-},\quad when \quad
\beta_{k}=0.\end{equation} Together with system (1.1) we consider
the differential system
\begin{equation} \frac{dY}{dz}=-Y(z)A(z),\quad z{\in}C,
\end{equation}where  $A(z)$ is defined by (1.2).The strong regular
solution of system (1.17) has the form (see \textbf{[Sa06]}):
\begin{equation}
Y(z)=\sum_{p{\geq}m}c_{p}(z-z_{k})^{p},\quad b_{m}{\ne}0.
\end{equation} The following relations
\begin{equation}
c_{q+1}[(q+1)I_{n}+a_{-1}]=-\sum_{j+\ell=q}c_{\ell}a_{j},
\end{equation}are true (see\textbf{[Sao6]}). Here $j{\geq}0,\quad
\ell{\geq}-1.$ Then we have \begin{equation}c_{-1}=P_{k}^{-},\quad
c_{0}=
-\sum_{j{\ne}k}P_{k}^{-}\frac{P_{j}}{(z_{k}-z_{j})}P_{k}.\end{equation}
From relations (1.19) and (1.20) we deduce that \begin{equation}
c_{1}P_{k}^{-}=P_{k}^{-}[\sum_{j{\ne}k}\frac{P_{j}}{z_{k}-z_{j}}\sum_{\ell{\ne}k}P_{k}\frac{P_{\ell}}{z_{k}-z_{\ell}}
+\sum_{j{\ne}k}\frac{P_{j}}{(z_{k}-z_{j})^{2}}].\end{equation}
 According condition (1.10) the matrix
 $P_{j}$ coincides with the transposed matrix
$P_{j}^{\tau}$, i.e.
\begin{equation}P_{j}=P_{j}^{\tau},\quad
1{\leq}j{\leq}s.\end{equation} By relations (1.6),(1.7), (1.22)
and the equalities
\begin{equation}P_{k}^{+}+P_{k}^{-}=2I_{n},\quad [P_{k}^{+}]^{2}+[P_{k}^{-}]^{2}=4I_{n}
\end{equation} we
obtain  that
\begin{equation}P_{k}^{-}[P_{j}P_{k}P_{\ell}+P_{\ell}P_{k}P_{j}]=[P_{j}P_{k}P_{\ell}+P_{\ell}P_{k}P_{j}]P_{k}^{-},\quad
j{\ne}k,\quad j{\ne}\ell,\quad k{\ne}\ell.\end{equation}
\begin{equation}P_{k}^{-}(P_{j}P_{k}P_{j}+P_{j})=(P_{j}P_{k}P_{j}+P_{j})P_{k}^{-},\quad
j{\ne}k\end{equation}\\
Hence  equation (1.21) has the solution
\begin{equation}
c_{1}=\sum_{j{\ne}k}\frac{P_{j}}{z_{k}-z_{j}}\sum_{\ell{\ne}k}P_{k}\frac{P_{\ell}}{z_{k}-z_{\ell}}
+\sum_{j{\ne}k}\frac{P_{j}}{(z_{k}-z_{j})^{2}},\quad when \quad
\beta_{k}{\ne}0\end{equation} and \begin{equation}
c_{1}=\sum_{j{\ne}k}\frac{P_{j}}{z_{k}-z_{j}}\sum_{\ell{\ne}k}P_{k}\frac{P_{\ell}}{z_{k}-z_{\ell}}
+\sum_{j{\ne}k}\frac{P_{j}}{(z_{k}-z_{j})^{2}}+
  P_{k}^{+},\quad when \quad \beta_{k}=0.\end{equation}It follows
from (1.1) and (1.17) that
\begin{equation}\frac{d}{dz}[W(z)Y(z)]=0.\end{equation}
Using  (1.4) and (1.18) we obtain
\begin{equation}W(z)Y(z)=b_{0}c_{0}+b_{-1}c_{1}+b_{1}c_{-1}.\end{equation}
Relations (1.12) ,(1.15),(1.16), and (1.20),(1.25),(1.26) imply
that
\begin{equation} b_{0}c_{0}=0,\quad
b_{-1}c_{1}+b_{1}c_{-1}=2\beta_{k}I_{n} \quad when \quad
\beta_{k}{\ne}0,\end{equation} and
\begin{equation}b_{-1}c_{1}+b_{1}c_{-1}=4I_{n} \quad when
\quad \beta_{k}=0,\end{equation} Hence we have
\begin{equation}
\mathrm{det}W(z)Y(z){\ne}0.\end{equation} In view of (1.32) the
constructed solutions $W(z)$ and $Y(z)$ of systems (1.1) and
(1.17) respectively are fundamental. It follows from (1.2) that
the point $z=\infty$ is the singular point of the first kind (see
\textbf{CL55, Ch.4}).As in our
 case the fundamental solutions $W(z)$ and $Y(z)$ are one-valued,
 then the following representations
\begin{equation} W(z)=\sum_{j=-\infty}^{m_{1}}g_{j}z^{j},\quad m_{1}<\infty,
 \quad |z|>R,\end{equation}
\begin{equation} Y(z)=\sum_{j=-\infty}^{m_{2}}h_{j}z^{j},\quad m_{2}<\infty,
 \quad |z|>R\end{equation} are true (see \textbf{CL55, Ch.4}).
 Here $ g_{j},\quad h_{j}$ are $n{\times}n$ matrices.Thus all the
 the points $z_{k}\quad (1{\leq}k{\leq}s)$ and $z=\infty$ are
 strong regular. Hence $W(z)$ and $Y(z)$ are rational matrix
 functions.We note that $Y^{\tau}(z)$ is the fundamental solution
 of system (1.1), when $\rho=-1.$
  The theorem is proved.\\
  \textbf{Remark 1.1.} When $s=1$ conditions 1) and 2) of Theorem
  1.1 must be omitted. When s=1 condition 2) of Theorem 1.1
  must be omitted.\\
\textbf{Remark 1.2.} Let the matrices $P_{j}$ be symmetric. If
system (1.1),(1.2) has a fundamental rational solution $W(z)$ ,
when $\rho=k$, then this system has the fundamental rational
solution $Y(z)=[W^{-1}(z)]^{\tau}$, when $\rho=-k.$\\
\textbf{Corollary 1.1.} \emph{Let conditions of Theorem $1.1$ be
fulfilled.Then the matrix functions $W(z)$ and $Y(z)$ can be
written in the forms}
\begin{equation}W(z)=\sum_{k=1}^{s}\frac{L_{k}}{z-z_{k}}+Q_{1}(z),\end{equation}
\begin{equation}Y(z)=\sum_{k=1}^{s}\frac{M_{k}}{z-z_{k}}+Q_{2}(z),\end{equation}
\emph{where $L_{k}$ and $M_{k}$ are $n{\times}n$ matrices,
$Q_{1}(z)$ and $Q_{2}(z)$ are  $n{\times}n$ matrix polynomials.}\\
Further we use the  relation
\begin{equation}A(z)=\frac{T}{z}[1+o(1)],\quad
z{\to}\infty,\end{equation} where
\begin{equation} T=\sum_{k=1}^{s}P_{k}. \end{equation}
Relation (1.37) and the strong regularity of the point $z=\infty$
imply the following assertion. \\
\textbf{Corollary 1.2.} \emph{All eigenvalues of the matrix $T$
are integer.}\\
We denote by $m_{T}$  the smallest eigenvalue  and by $M_{T}$ the
greatest eigenvalue of $T$. Changing the variable $z=\frac{1}{u}$
in system (1.1) we obtain the following results.\\
\textbf{Corollary 1.3.} \emph{Let matrix polynomials $Q_{1}(z)$
and $Q_{2}(z)$ be defined by relations} (1.35) \emph{and}
(1.36).\\
1.  \emph{If $M_{T}{\geq}0$ then $\mathrm{deg} Q_{1}(z)=M_{T}$}.\\
2. \emph{If $M_{T}<0$ then $Q_{1}(z)=0.$}\\
3. \emph{If $m_{T}{\leq}0$ then $\mathrm{deg} Q_{2}(z)=-m_{T}$}.\\
4. \emph{If $m_{T}>0$ then $Q_{2}(z)=0$}.\\
\section{Representation of the symmetric group $S_{n}$}
Let $S_{n}$ be the symmetric group. We consider the natural
representation of $S_{n}$. By $(i;j)$ we denote the permutation
which transposes $i$ and $j$ and preserves all the rest. The
 $n{\times}n$ matrix which corresponds to $(i;j)$ is denoted by
 \begin{equation}
 P(i,j)=[p_{k,\ell}(i,j)], \quad (i{\ne}j).\end{equation}
 The elements $p_{k,\ell}(i,j)$ are equal to zero except the
 following cases
 \begin{equation}p_{k,\ell}(i,j)=1,\quad (k=i,\ell=j);\quad p_{k,\ell}(i,j)=1,\quad
(k=j,\ell=i),\end{equation}
 \begin{equation}p_{k,k}(i,j)=1,\quad
 (k{\ne}i,k{\ne}j).\end{equation} Now we introduce the matrices
 \begin{equation}
 P_{k}=P(1,k+1),
 \quad 1{\leq}k{\leq}n-1.\end{equation}
 \textbf{Proposition 2.1.} \emph{The matrices $P_{k}, \quad (1{\leq}k{\leq}n-1)$
 satisfy the conditions 1)-4) of Theorem 1.1.}\\
 \emph{Proof.} It follows from relations (2.1) - (2.4) that conditions
 1) and 4) are fulfilled. By direct calculation we see that
 \begin{equation}[P_{1}P_{2}P_{1}+P_{1}]P_{2}^{+}=P_{2}^{+},\quad
 n=3,\end{equation}
 \begin{equation}[P_{1}P_{2}P_{3}+P_{3}P_{2}P_{1}]P_{2}^{+}=0,\quad
 n=4.\end{equation} Hence conditions 2) and 3) are fulfilled for
 all $j,k,\ell$ and $n$.The proposition is proved.\\
 \textbf{Corollary 2.1.} \emph{The system $(1.1)$, $(1.2)$ has a
 rational fundamental solution, when} $\rho={\pm}1,\quad
 P_{k}=P(1,k+1).$\\
 Corollary 2.1.confirms the conjecture of A.Chervov and D.Talalaev
 (see\textbf{[CT06]} for the case $\rho={\pm}1.$\\
 \textbf{Remark 2.1.} The natural representation of $S_{n}$ is the
 sum of the 1-reperesentation and an irreducible representation
  (see \textbf[Bu65, p. 106]).\\
 We introduce the $1{\times}(n-1)$ vector $e=[1,1, ...,1]$ and
 $n{\times}n$ matrix \begin{equation}
 T_{1}=\left[\begin{array}{cc}
   2-n & e\\
   e^{\tau} & 0 \\
 \end{array}\right].
 \end{equation}
 Using relations (1.38) and (2.1)-(2.4) we deduce that
 \begin{equation}
 T=(n-2)I_{n}+T_{1}.\end{equation}
 The eigenvalues of $T$ are defined by the equalities
 \begin{equation} \lambda_{1}=n-1,\quad \lambda_{2}=n-1,\quad
 \lambda_{3}=-1.\end{equation}Hence we have $m_{T}=-1,\quad
 M_{T}=n-1.$ It follows from Corollary 1.3 the statement.\\
 \textbf{Proposition 2.2.} \emph{ Let the matrices $P_{k}$ are
 defined by relations $(2.1)-(2.4)$. Then the equalities
 \begin{equation} \mathrm{deg} Q_{1}(z)=n-1,\quad \mathrm{deg} Q_{2}(z)=-1 \end
 {equation} are true.}\\
 \emph{Acnowledgement.} I would like to express my gratitude
 to A.Chervov for the useful discussion.
 \begin{center} References \end{center}
 \textbf{[Bu65]} Burrow M., Representation Theory of Finite Groups, Academic
 Press, 1965.\\
 \textbf{[ST06]} Chervov A., Talalaev D.,Quantum Spectral
 Curves, Quantum Integrable Systems and the
 Geometric Langlands Correspondence, arXiv:hep-th/0604128, 2006.\\
 \textbf{[CL55]} Coddington E.A., Levinson N., Theory of Ordinary
 Differential Equations, McGraw-Hill Company, New York, 1955.\\
 \textbf{[Sa06]} Sakhnovich L.A., Meromorphic Solutions of Linear
 Differential Systems, Painleve Type Functions, arxiv: math.CA/0607555, 2006.
\end{document}